\documentclass[a4paper]{jpconf}
\usepackage{graphicx}

\newcommand {\beq}{\begin{equation}}
\newcommand {\eeq}{\end{equation}}
\newcommand {\bea}{\begin{eqnarray}}
\newcommand {\eea}{\end{eqnarray}}
\newcommand {\nn}{\nonumber \\}
\newcommand {\m}{\mu}
\newcommand {\n}{\nu}
\newcommand {\pl}{\partial}

\newcommand {\al}{\alpha}
\newcommand {\be}{\beta}

\newcommand {\la}{\lambda}
\newcommand {\La}{\Lambda}

\newcommand {\om}{\omega}

\newcommand {\ep}{\epsilon}

\newcommand {\na}{\nabla}
\newcommand {\del}  {\delta}
\newcommand {\Del}  {\Delta}
\newcommand {\mn}{{\mu\nu}}

\newcommand {\half}{ {\frac{1}{2}} }

\newcommand {\fourth} {\frac{1}{4} }

\newcommand {\Ecal}{{\cal E}}
\newcommand {\Fcal}{{\cal F}}
\newcommand {\Lcal}{{\cal L}}

\newcommand {\Pcal}{{\cal P}}
\newcommand {\Dcal}{{\cal D}}

\newcommand {\ptil} {{\tilde p}}
\newcommand {\ktil} {{\tilde k}}

\newcommand {\omtil}{{\tilde \omega}}

\newcommand {\Lhat}{{\hat L}}

\newcommand {\delh} {{\hat \delta}}

\newcommand {\bfZ} {{\bf Z}}
\newcommand {\K}{{\bf K}}
\newcommand {\I}{{\bf I}}

\newcommand {\change} {\leftrightarrow}
\newcommand {\ra} {\rightarrow}

\newcommand {\pr}   {{\quad .}}
\newcommand {\com}  {{\quad ,}}
\newcommand {\q}    {\quad}

\newcommand {\PTP}  {{\it Prog.Theor.Phys.}}

\newcommand {\intxz} {{\int d^4xdz}}

\newcommand {\intp} {{\int \frac{d^4p}{(2\pi)^4}}}
\newcommand {\intpE} {{\int \frac{d^4p_E}{(2\pi)^4}}}

\newcommand {\intt} {{\int_{0}^{\infty}\frac{dt}{t}}}
\newcommand {\Pla} {\frac{{\tilde p}}{\omega}}

\newcommand {\Tev} {\frac{{\tilde p}}{T}}

\begin{document}
\title{Casimir Energy of the Universe and New Regularization of Higher Dimensional Quantum Field Theories}

\author{Shoichi Ichinose}

\address{
Laboratory of Physics, School of Food and Nutritional Sciences, 
University of Shizuoka, 
Yada 52-1, Shizuoka 422-8526, Japan}

\ead{ichinose@u-shizuoka-ken.ac.jp}

\begin{abstract}
Casimir energy is calculated for the 5D electromagnetism and 5D scalar theory
in the {\it warped} geometry. 
It is compared with the flat case. 
A new regularization, 
called {\it sphere lattice regularization}, is taken. 
In the integration over the 5D space, we introduce two boundary 
curves (IR-surface and UV-surface) based on the {\it minimal area principle}. 
It is a {\it direct} realization of the geometrical approach 
to the {\it renormalization group}.  
The regularized configuration is {\it closed-string like}. 
We do {\it not} take the KK-expansion approach. Instead, 
the position/momentum propagator is exploited, 
combined with the {\it heat-kernel method}. All expressions
are closed-form (not KK-expanded form). 
The {\it generalized} P/M propagators are introduced. 
We numerically evaluate $\La$(4D UV-cutoff), $\om$(5D bulk curvature, 
warp parameter)
and $T$(extra space IR parameter) dependence of the Casimir energy. We present 
two {\it new ideas} in order to define the 5D QFT:\  
1) the summation (integral) region over the 5D space is {\it restricted} by two minimal surfaces 
(IR-surface, UV-surface) ; or 
2) we introduce a {\it weight function} and require the dominant contribution, in the summation, 
is given by the {\it minimal surface}. 
Based on these, 
5D Casimir energy is {\it finitely} obtained after the {\it proper renormalization
procedure.} 
The {\it warp parameter} $\om$ suffers from the {\it renormalization effect}. 
The IR parameter $T$ does not. 
We examine the meaning of the weight function and finally 
reach a {\it new definition} of the Casimir energy where {\it the 4D momenta( or coordinates) 
are quantized} with the extra coordinate as the Euclidean time (inverse temperature). 
We examine the cosmological constant problem and present an answer at the end. 
Dirac's large number naturally appears. 
\end{abstract}

\section{Introduction}
In the dawn of the quantum theory, the {\it divergence} problem of
the specific heat of the radiation cavity was the biggest one 
(the problem of the blackbody radiation). It is historically so famous 
that the difficulty was solved by Planck's idea that the energy is quantized. 
In other words, the phase space of the photon field dynamics is not continuous but 
has the "cell" or "lattice" structure with the unit area ($\Del x\cdot\Del p$) of the size $2\pi\hbar$ (Planck constant). 
The radiation energy is composed of two parts, $E_{Cas}$ and $E_\be$:
\bea
E_{4dEM}=E_{Cas}+E_\be\ ,\ 
E_{Cas}=\sum_{m_x,m_y,n\in\bfZ}\omtil_{m_xm_yn}\ ,\ 
E_{\be}=2\sum_{m_x,m_y,n\in\bfZ}\frac{\omtil_{m_xm_yn}}{\e^{\be\omtil_{m_xm_yn}}-1},\nn
{\omtil_{m_xm_yn}}^2=(m_x\frac{\pi}{L})^2+(m_y\frac{\pi}{L})^2+(n\frac{\pi}{l})^2
\com\q\q l\ll L\com
\label{4dEM18X}
\eea 
where the parameter $\be$ is the inverse temperature, $l$ is the separation length 
between two perfectly-conducting plates, and $L$ is the IR regularization parameter 
of the plate-size.  
The second part $E_\be$ 
is, essentially, Planck's radiation formula. The first one $E_{Cas}$ is 
the vaccuum energy of the radiation field, that is, the Casimir energy. 
It is a very delicate quantity. The quantity is formally {\it divergent}, hence it 
must be defined with careful {\it regularization}. $E_{Cas}/(2L)^2$ does 
depend only on the {\it boudary} parameter $l$. The quantity is a quantum effect and 
, at the same time, depends on the {\it global} (macro) parameter $l$. 
\bea
\frac{E_{Cas}}{(2L)^2}=\frac{\pi^2}{(2l)^3}\frac{B_4}{4!}=-\frac{\pi^2}{720}\frac{1}{(2l)^3}\com\q
B_4\mbox{(the fourth Bernoulli number)}=-\frac{1}{30}
\com
\label{4dEM24x}
\eea 

\begin{figure}[h]
\begin{minipage}{18pc}
\includegraphics[width=18pc]{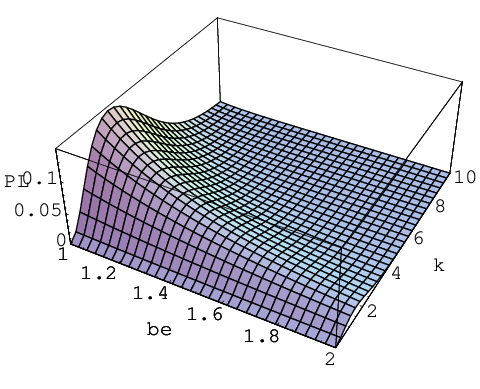}
\caption{\label{PlanckDistB}Graph of Planck's radiation formula.  
$ \Pcal (\be,k)=\frac{1}{(c\hbar)^3}\frac{1}{\pi^2}k^3/(\e^{\be k}-1)\ \ 
(1\leq\be\leq 2,\ 0.01\leq k\leq 10)$.}
\end{minipage}\hspace{2pc}%
\begin{minipage}{18pc}
\includegraphics[width=18pc]{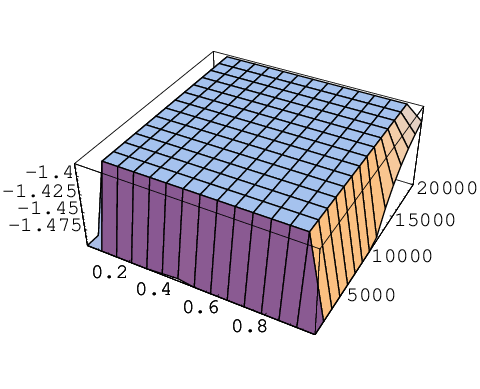}
\caption{\label{FcalmHT1k10p4}Behavior of $\ln |\half\Fcal^-(\ktil,z)|=\ln |\ktil~ G^-_k(z,z)/(\om z)^3|$. 
$\om=10^4, T=1, \La=2\times 10^4$. $1.0001/\om \leq z \leq 0.9999/T$. $\La T/\om \leq \ktil \leq \La$. 
Note $\ln |(1/2)\times (1/2)|\approx -1.39$.}
\end{minipage} 
\end{figure}

In Fig.\ref{PlanckDistB}, Planck's radiation spectrum distribution is shown. 
Introducing the axis of the inverse temperature($\be$), besides the photon energy or 
frequency ($k$), 
it is shown stereographically. Although we will examine the 5D version of 
the zero-point part (the Casimir energy), the calculated quantities in this paper 
are much more related to this Planck's formula. 
We see, near the $\be$-axis, a sharply-rising surface, which is the Rayleigh-Jeans 
region (the energy density is proportional to the {\it square} of the photon frequency). 
The {\it damping} region in high $k$ is the Wien's region.
\footnote{
We recall that the old problem of the {\it divergent} specific heat was 
solved by the Wien's formula. This fact strongly supports 
the present idea of introducing the {\it weight function} 
(see Sec.\ref{uncert}). 
} 
The ridge (the line of peaks at each $\be$) 
forms the {\it hyperbolic} curve (Wien's displacement law). When we will, in this paper, 
deal with the 
energy distribution over the 4D momentum and the extra-coordinate, we will see the similar 
behavior (although top and bottom appear in the opposite way).  

In the quest for the unified theory, the higher dimensional (HD) approach 
is a fascinating one from the geometrical point. Historically the initial successful 
one is the Kaluza-Klein model\cite{Kal21,Klein26}, which unifies the photon, graviton and dilaton
from the 5D space-time approach. 
The HD theories
, however, generally have the serious defect as the quantum field
theory(QFT) : un-renormalizability. 
The HD quantum field theories, at present, 
are not defined within the QFT. One can take the standpoint 
that the more fundamental formulation, such as the string theory and D-brane 
theory, can solve the problem. In the present paper, we have the {\it new standpoint} 
that the HD theories should be defined by themselves within the QFT.  In order to 
escape the dimension requirement D=10 or 26 
from the quantum consistency (anomaly cancellation)\cite{StringText}, 
we treat the gravitational (metric) field only as the background one. 
This does {\it not} mean the space-time
is not quantized. See later discussions (Sec.\ref{weight}). We present a way to define 5D quantum field theory 
through the analysis of the Casimir energy of 5D electromagnetism. 

In 1983, the Casimir energy in the Kaluza-Klein theory was calculated 
by Appelquist and Chodos\cite{AC83}. They took the cut-off ($\La$) regularization and 
found the quintic ($\La^5$) divergence and the finite term. The divergent 
term shows the {\it unrenormalizability} of the 5D theory, but the finite term looks 
meaningful
\footnote{
The gauge independence was confirmed in Ref.\cite{SI85PLB}. 
}
 and, in fact, is widely regarded as the right vacuum energy 
which shows {\it contraction} of the extra axis. 

In the development of the string and D-brane theories, a new approach 
to the renormalization group was found. It is called {\it holographic renormalization} 
\cite{SusWit9805,HenSken9806,SkenTown9909,DFGK9909,FGPW9904,dBVV9912}. 
We regard the renormalization flow as a curve 
in the bulk (HD space). The flow goes along the extra axis. 
The curve is derived as a dynamical equation 
such as Hamilton-Jacobi equation. 
It originated from the AdS/CFT correspondence\cite{Malda9711,GKP9802,Witten9802}. 
Spiritually the present basic idea overlaps with this approach. 
The characteristic points of this paper are:\  
a) We do {\it not} rely on the 5D supergravity;\  
b) We do {\it not} quantize the gravitational(metric) field;\ 
c) The divergence problem is solved by reducing the degree of freedom of the system, 
where we require, not higher symmetries, but some restriction based on the {\it minimal area principle};\ 
d) No local counterterms are necessary.  @
   
In the previous paper\cite{SI0801}, we investigated the 5D electromagnetism 
in the {\it flat} geometry. 
The results show the {\it renormalization of the compactification size} $l$. 
\bea
E^W_{Cas}/\La l =-\frac{\al}{l^4}\left( 1-4c\ln (l\La) \right) =-\frac{\al}{{l'}^4}\com\q
\be=\frac{\pl}{\pl (\ln \La)}\ln\frac{l'}{l}=c\com
\label{uncert1cc}
\eea
where $\al$ and $c$ are some numbers. They are, at present, not fixed, but
are numerically obtained depending on the weight function $W$. 
The aim of this paper is to examine how the above results change 
for the 5D warped geometry case. 
One additional massive parameter, that is, 
the warp (bulk curvature) parameter $\om$ appears. 
This introduction of the "thickness" $1/\om$ 
comes from the expectation that it softens the UV-singularity, which 
is the same situation as in the string theory.

\section{Kaluza-Klein expansion approach\label{KKexp}}
In order to analyze the 5D EM-theory, we start with 5D {\it massive} vector
theory. 
\bea
S_{5dV}=\intxz\sqrt{-G}(-\fourth F_{MN}F^{MN}-\half m^2A^M A_M)\ ,\ 
F_{MN}=\pl_M A_N-\pl_NA_M\ ,\nn
ds^2=\frac{1}{\om^2z^2}(\eta_\mn dx^\m dx^\n+{dz}^2)=G_{MN}dX^M dX^N\com\q G\equiv \det G_{AB}
\pr
\label{KKexp1}
\eea 
The 5D vector mass, $m$, is regarded as a IR-regularization parameter. 
In the limit, $m=0$, the above one has the 5D local-gauge symmetry. 
Casimir energy is given by some integral where the (modified) Bessel functions, 
with the index $\nu=\sqrt{1+\frac{m^2}{\om^2}}$, appear. 
Hence the 5D EM 
limit is given by $\n=1\ (m=0)$. We consider, however, the {\it imaginary} 
mass case $m=i\om\ (m^2=-\om^2,\ \n=0)$ mainly for the simplicity.  
We can simplify the model furthermore. 
Instead of analyzing the $m^2=-\om^2$ of the massive vector (\ref{KKexp1}), 
we take the 5D massive {\it scalar} theory
on AdS$_5$ with $m^2=-4\om^2,\ \n=\sqrt{4+m^2/\om^2}=0$. 
\bea
\Lcal=\sqrt{-G}(-\half \na^A\Phi\na_A\Phi-\half m^2\Phi^2),\ 
ds^2=G_{AB}dX^AdX^B,\ \na^A\na_A \Phi-m^2\Phi+J=0,
\label{KKexp1b}
\eea 
where $\Phi(X)=\Phi(x^a,z)$ is the 5D scalar field. 
The integral region is given by 
\bea
-\frac{1}{T}\leq z \leq -\frac{1}{\om}\q\mbox{or}\q
\frac{1}{\om}\leq z \leq \frac{1}{T}\q 
(-l\leq y \leq l\ ,\ |z|=\frac{1}{\om}\e^{\om |y|})\com\q
\frac{1}{T}\equiv \frac{1}{\om}\e^{\om l}
\com
\label{KKexp2}
\eea 
where we take into account $Z_2$ symmetry: $z\change -z$.  
$\om$ is the bulk curvature (AdS$_5$ parameter) and $T^{-1}$ is the
size of the extra space (Infrared parameter). 
In this section, we present the standard expressions, that is, 
those obtained by the Kaluza-Klein expansion. 
The Casimir energy $E_{Cas}$ is given by
\bea
\e^{-T^{-4}E_{Cas}}=\int\Dcal\Phi\exp\{i\int d^5X\Lcal\}\nn
=\int\Dcal\Phi_p(z)\exp\left[
i\intp 2\int_{1/\om}^{1/T}dz\left\{
\half\Phi_p(z)s(z)( {s(z)}^{-1}\Lhat_z-p^2)\Phi_p(z)
                             \right\} 
                          \right]\nn
=\int\prod_n dc_n(p)\exp\left[
\intpE\sum_n\{
-\half c_n(p)^2(p_E^2+M_n^2)
             \}
                         \right]
=\exp\sum_{n,p}\{-\half\ln (p_E^2+M_n^2) \}
\com
\label{KKexp3}
\eea 
where $\Phi_p(z)$ is the partially(4D world only)-Fourier-transformed 
one of $\Phi(X)$. $\Phi_p(z)$ is expressed in the expansion form 
using the eigen functions, $\psi_n(z)$, of this AdS$_5$ system. 
\bea
\Phi_p(z)=\sum_{n}c_n(p)\psi_n(z), 
\{s(z)^{-1}\Lhat_z+{M_n}^2\}\psi_n(z)=0, 
\Lhat_z\equiv\frac{d}{dz}\frac{1}{(\om z)^3}\frac{d}{dz}
                           -\frac{m^2}{(\om z)^5},\nn
\psi_n(z)=-\psi_n(-z)\q\mbox{for}\q P=-\q ;\q
\psi_n(z)=\psi_n(-z)\q\mbox{for}\q P=+ 
\pr
\label{KKexp8}
\eea
where $s(z)=\frac{1}{(\om z)^3}$. 
The expression (\ref{KKexp3}) is the familiar one 
of the Casimir energy.

\section{Heat-Kernel Approach and Position/Momentum Propagator\label{HKA}}
Eq.(\ref{KKexp3}) is the expression of $E_{Cas}$ by the KK-expansion.
In this section, the same quantity is re-expressed in a {\it closed} form
using the heat-kernel method and the P/M propagator. 
First we can express it, using the heat equation solution, as follows. 
\bea
\e^{-T^{-4} E_{Cas}}
=(\mbox{const})\times\exp \left[ 
T^{-4}\intp 2\int_{0}^{\infty}\half\frac{dt}{t}\mbox{Tr}~H_{p}(z,z';t) 
                          \right] \com\nn
\mbox{Tr}~H_p(z,z';t)=\int_{1/\om}^{1/T}s(z)H_p(z,z;t)dz\com\q
\{\frac{\pl}{\pl t}-(s^{-1}\Lhat_z-p^2) \}H_p(z,z';t)=0
\pr
\label{HKA2}
\eea
The heat kernel $H_p(z,z';t)$ is formally solved, using the
Dirac's bra and ket vectors $(z|, |z)$, as
\bea
H_p(z,z';t)=(z|\e^{-(-s^{-1}\Lhat_z+p^2)t}|z')
\pr
\label{HKA3}
\eea
We here introduce the position/momentum propagators $G^{\mp}_p$ as
follows.
\bea
G^\mp_p(z,z')\equiv
\int_0^\infty dt~ H_p(z,z';t)=
\sum_{n\in \bfZ}\frac{1}{M_n^2+p^2}
\half\{ \psi_n(z)\psi_n(z')\mp \psi_n(z)\psi_n(-z') \}
\pr
\label{HKA6}
\eea
They satisfy the following differential equations of {\it propagators}.
\bea
(\Lhat_z-p^2s(z))G^{\mp}_p(z,z')=
\sum_{n\in \bfZ}
\frac{
\{ \psi_n(z)\psi_n(z')\mp \psi_n(z)\psi_n(-z') \}
     }{2}
=\left\{
\begin{array}{ll}
\ep(z)\ep(z')\delh (|z|-|z'|) & \mbox{\ P=}-1 \\
\delh (|z|-|z'|) & \mbox{\ P=}1 
\end{array}
        \right.
\label{HKA7}
\eea

Therefore the Casimir energy $E_{Cas}$ is given by
\bea
-E^{-}_{Cas}(\om,T)=
\intpE 2\intt 2\int_{1/\om}^{1/T} dz~s(z)H_{p_E}(z,z;t)\nn
=\intpE 2\intt 2\int_{1/\om}^{1/T} dz~s(z)\left\{
\sum_{n\in \bfZ}\e^{-(M_n^2+p_E^2)t}\psi_n(z)^2
                                       \right\}
\com
\label{HKA8}
\eea
where $s(z)=1/(\om z)^3$. 
The momentum symbol $p_E$ indicates Euclideanization. 
This expression leads to the same treatment as the previous section. 
Note that the above expression shows the {\it negative definiteness} of 
$E^{-}_{Cas}$. 
Finally we obtain the following useful expression of the Casimir energy for $P=\mp$.
\bea
-E^\mp_{Cas}(\om,T)
=\intpE\int_{1/\om}^{1/T}dz~ s(z)\int_{p_E^2}^\infty\{G_k^\mp(z,z) \}dk^2
\pr
\label{HKA11}
\eea

The P/M propagators $G_p^\mp$ in (\ref{HKA6}) and (\ref{HKA11}) can be expressed in a {\it closed} form.
Taking the {\it Dirichlet} condition at all fixed points, the expression
for the fundamental region ($1/\om \leq z\leq z'\leq 1/T$) is given by
\bea
G_p^\mp(z,z')=\mp\frac{\om^3}{2}z^2{z'}^2
\frac{\{\I_0(\Pla)\K_0(\ptil z)\mp\K_0(\Pla)\I_0(\ptil z)\}  
      \{\I_0(\Tev)\K_0(\ptil z')\mp\K_0(\Tev)\I_0(\ptil z')\}
     }{\I_0(\Tev)\K_0(\Pla)-\K_0(\Tev)\I_0(\Pla)},
\label{HKA12}
\eea 
where $\ptil\equiv\sqrt{p^2},p^2\geq 0$. 
We can express Casimir energy in terms of the following functions $F^\mp(\ptil,z)$. 
\bea
-E^{\La,\mp}_{Cas}(\om,T)
=\left.\intpE\right|_{\ptil\leq\La}\int_{1/\om}^{1/T}dz~F^\mp(\ptil,z) \com\nn
F^\mp(\ptil,z)\equiv s(z)\int_{p_E^2}^{\La^2}\{G_k^\mp(z,z) \}dk^2
=\frac{2}{(\om z)^3}\int_\ptil^\La\ktil~ G^\mp_k(z,z)d\ktil
\equiv \int_\ptil^\La\Fcal^\mp(\ktil,z)d\ktil
\com
\label{HKA13}
\eea
where $\Fcal^\mp(\ktil,z)$ are the integrands of $F^\mp(\ptil,z)$ and $\ptil=\sqrt{p_E^2}$. 
Here we introduce the UV cut-off parameter $\La$ for the 4D momentum space. 
In Fig.\ref{FcalmHT1k10p4}, we show the behavior of $\Fcal^-(\ktil,z)$. 
The table-shape graph says the "Rayleigh-Jeans" dominance.
That is, for the wide-range region $(\ptil,z)$ satisfying both 
$\ptil (z-\frac{1}{\om})\gg 1$ and $\ptil (\frac{1}{T}-z)\gg 1$, 
\bea
\Fcal^-(\ptil,z)\approx \half\com\q
\Fcal^+(\ptil,z)\approx \half
\com\q
(\ptil,z)\in \{(\ptil,z)| \ptil (z-\frac{1}{\om})\gg 1\ \mbox{and}\ \ptil (\frac{1}{T}-z)\gg 1\}
\pr
\label{HKA14}
\eea

\section{UV and IR Regularization Parameters and Evaluation of Casimir Energy
\label{UIreg}}

\begin{figure}[h]
\begin{minipage}{18pc}
\includegraphics[width=18pc]{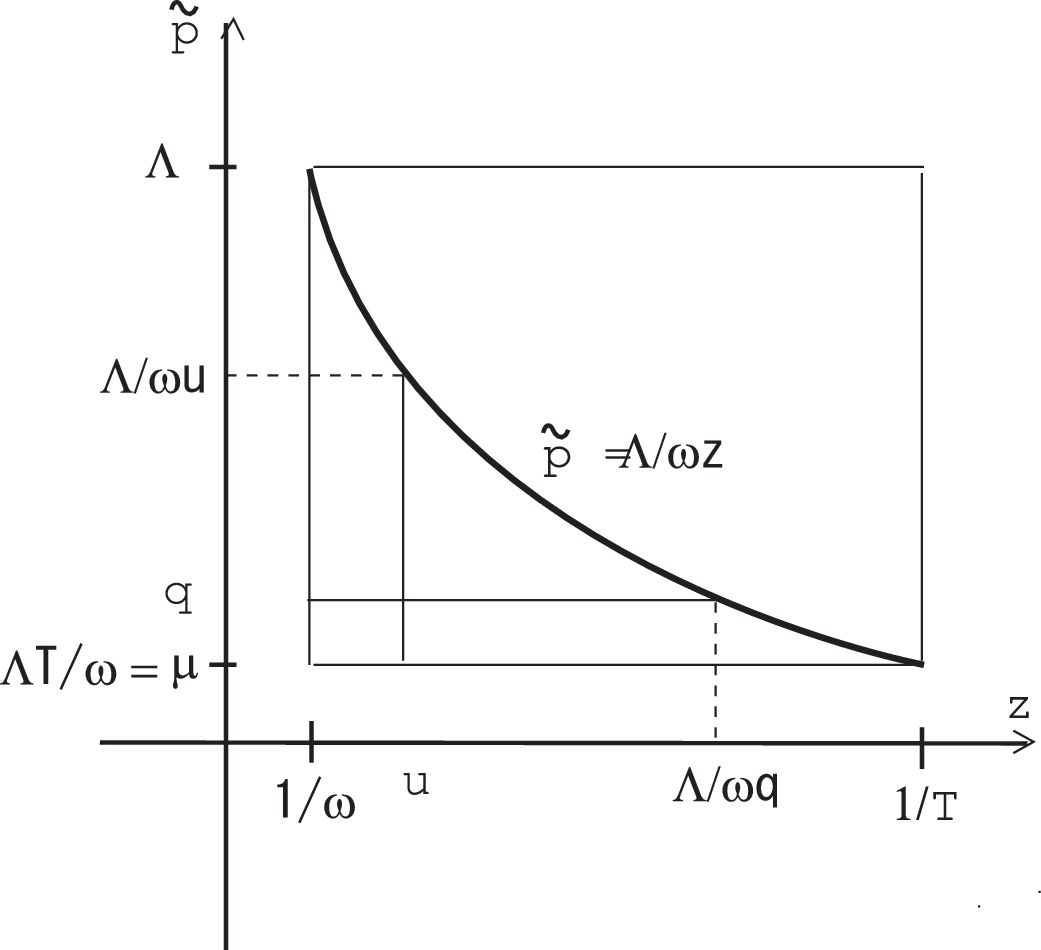}
\caption{\label{zpINTregionW}Space of (z,$\ptil$) for the integration. The hyperbolic curve 
will be used in Sec.\ref{surf}.}
\end{minipage}\hspace{2pc}%
\begin{minipage}{18pc}
\includegraphics[width=18pc]{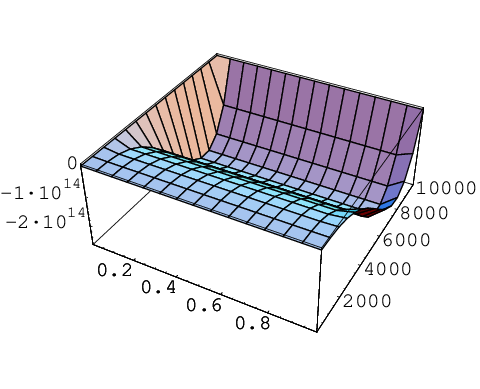}
\caption{\label{p3FmL10000}Behaviour of $(-1/2)\ptil^3F^-(\ptil,z)$ (\ref{UIreg2b}). $T=1, \om=10^4, \La=10^4$.  
$1.0001/\om\leq z<0.9999/T$, $\La T/\om\leq\ptil\leq \La$.}
\end{minipage} 
\end{figure}

The integral region of the above equation (\ref{HKA13}) is displayed in Fig.\ref{zpINTregionW}. 
In the figure, we introduce the regularization cut-offs for the 4D-momentum integral, 
$\m\leq\ptil\leq\La$. 
As for the extra-coordinate integral, it is the finite interval, 
$1/\om\leq z\leq 1/T=\e^{\om l}/\om$, hence we need not introduce further
regularization parameters. 
For simplicity, we take
the following IR cutoff of 4D momentum\ :\ 
$\m=\La\cdot\frac{T}{\om}=\La \e^{-\om l}$ . 
Hence the new regularization parameter is $\La$ only. 

Let us evaluate the ($\La,T$)-{\it regularized} value of (\ref{HKA13}). 
\bea
-E_{Cas}^{\La,\mp}(\om,T)=\frac{2\pi^2}{(2\pi)^4}
\int_{\m}^{\La}d\ptil\int_{1/\om}^{1/T}dz~\ptil^3 F^\mp (\ptil,z)\com\q 
F^\mp (\ptil,z)
=\frac{2}{(\om z)^3}\int_\ptil^\La\ktil~ G^\mp_k(z,z)d\ktil
\pr
\label{UIreg2b}
\eea
The integral region of ($\ptil,z$) is the {\it rectangle} shown in Fig.\ref{zpINTregionW}. 
Note that eq.(\ref{UIreg2b}) is the {\it rigorous} expression of the $(\La,T)$-regularized Casimir energy. 
We show the behavior of $(-1/2)\ptil^3F^-(\ptil,z)$ taking 
the values $\om=10^4, T=1$ in Fig.\ref{p3FmL10000}($\La=10^4$).
\footnote{
The requirement for the three parameters $\om, T, \La$ is $\La\gg\om\gg T$. 
See ref.\cite{SI01CQG} for the discussion about the hierarchy $\La, \om, T$.
}
Behavior
along $\ptil$-axis does not so much depend on $z$. A valley runs parallel to the $z$-axis 
with the bottom line 
at the fixed ratio of $\ptil/\La \sim 0.75$. 
The depth of the valley is proportional to $\La^4$. 
Because $E_{Cas}$ is the ($\ptil,z$) 'flat-plane' integral of $\ptil^3F(\ptil,z)$ , the {\it volume}
inside the valley is the quantity $E_{Cas}$ . Hence it is easy to see $E_{Cas}$ is proportional
to $\La^5$. This is the same situation as the flat case. 
Importantly, (\ref{UIreg2b}) shows the {\it scaling} behavior for large values of $\La$ and $1/T$. 
From a {\it close} numerical analysis of ($\ptil,z$)-integral (\ref{UIreg2b}), 
we have confirmed\ :\ Eq.(4A)\ 
$E^{\La,-}_{Cas}(\om,T)=\frac{2\pi^2}{(2\pi)^4}\times\left[ -0.0250 \frac{\La^5}{T} \right]$. 
It does {\it not} depend on $\om$ and has no $\ln \frac{\La}{T}$-term. 
(Note: $0.025=1/40$.) 
Compared with the flat case, we see the factor $T^{-1}$ 
plays the role of {\it IR parameter} of the
extra space. We note that the behavior of Fig.\ref{p3FmL10000} 
is similar to the Rayleigh-Jeans's region (small momentum region) of the Planck's radiation 
formula (Fig.\ref{PlanckDistB}) in the sense that 
$\ptil^3F(\ptil,z)\propto \ptil^3$ for $\ptil\ll\La$. 

Finally we notice, from the Fig.\ref{p3FmL10000}, the approximate form 
of $F(\ptil,z)$ for the large $\La$ and $1/T$ is given by\ :\ Eq.(4B)\ 
$F^\mp (\ptil,z)\approx \frac{f}{2} \La (1-\frac{\ptil}{\La}),\ f=1$.
It does {\it not} depend on $z, \om$ and $T$. $f$ is the degree of freedom. 
The above result is consistent with (\ref{HKA14}).

\section{UV and IR Regularization Surfaces, Principle of 
Minimal Area and Renormalization Flow\label{surf}}
The advantage of the
new approach is that the KK-expansion is replaced by the integral 
of the extra dimensional coordinate $z$ and 
all expressions are written in the {\it closed} (not expanded)
form. The $\La^5$-divergence, (4A), shows the notorious problem
of the higher dimensional theories, as in the flat case. 
In spite of all efforts of the past literature, 
we have not succeeded 
in defining the higher-dimensional theories. 
(The divergence causes problems. The famous example is 
the divergent {\it cosmological constant} in the gravity-involving theories.
\cite{AC83} )
Here we notice that the divergence problem can be solved if we find a way to
{\it legitimately restrict the integral region in ($\ptil,z$)-space}. 

One proposal of this was presented by Randall and Schwartz\cite{RS01}. They introduced
the {\it position-dependent cut-off},\ $\mu <\ptil <\La /\om u\ ,\ u\in [1/\om,1/T]$\ , 
for the 4D-momentum integral in the "brane" located at $z=u$. See Fig.\ref{zpINTregionW}.
The total integral region is the lower part of the {\it hyperbolic} curve $\ptil=\La /\om z$. 
They succeeded in obtaining the {\it finite} $\be$-function of the 5D warped vector
model. 
We have confirmed that the value $E_{Cas}$ of (\ref{UIreg2b}), when the Randall-Schwartz 
integral region (Fig.\ref{zpINTregionW}) is taken, is proportional to $\La^5$. 
The close numerical analysis says
\bea
E^{-RS}_{Cas}(\om,T)=
\frac{2\pi^2}{(2\pi)^4}\int_{\m}^{\La}dq\int_{1/\om}^{\La /\om q}dz~q^3 F^- (q,z)
=\frac{2\pi^2}{(2\pi)^4}\int_{1/\om}^{1/T}du\int_{\m}^{\La /\om u}d\ptil~\ptil^3 F^- (\ptil,u)\nn
=\frac{2\pi^2}{(2\pi)^4}\frac{\La^5}{\om}\left\{
-1.58\times 10^{-2}-1.69\times 10^{-4}\ln~\frac{\La}{\om}
                                          \right\}
\com
\label{surfM1}
\eea
which is {\it independent} of $T$ 
.
This shows the divergence 
situation does {\it not} improve compared with the non-restricted case of (4A). 
{\it $T$ of (4A) is replaced by the warp parameter $\om$.} 
This is contrasting with the flat case where $E^{RS}_{Cas}\propto -\La^4$. 
The UV-behavior, however, {\it does improve} if we can choose the parameter $\La$ in the way: $\La\propto\om$.
This fact shows the parameter $\om$ "smoothes" the UV-singularity to some extent. 

\begin{figure}[h]
\begin{minipage}{22pc}
\includegraphics[width=22pc]{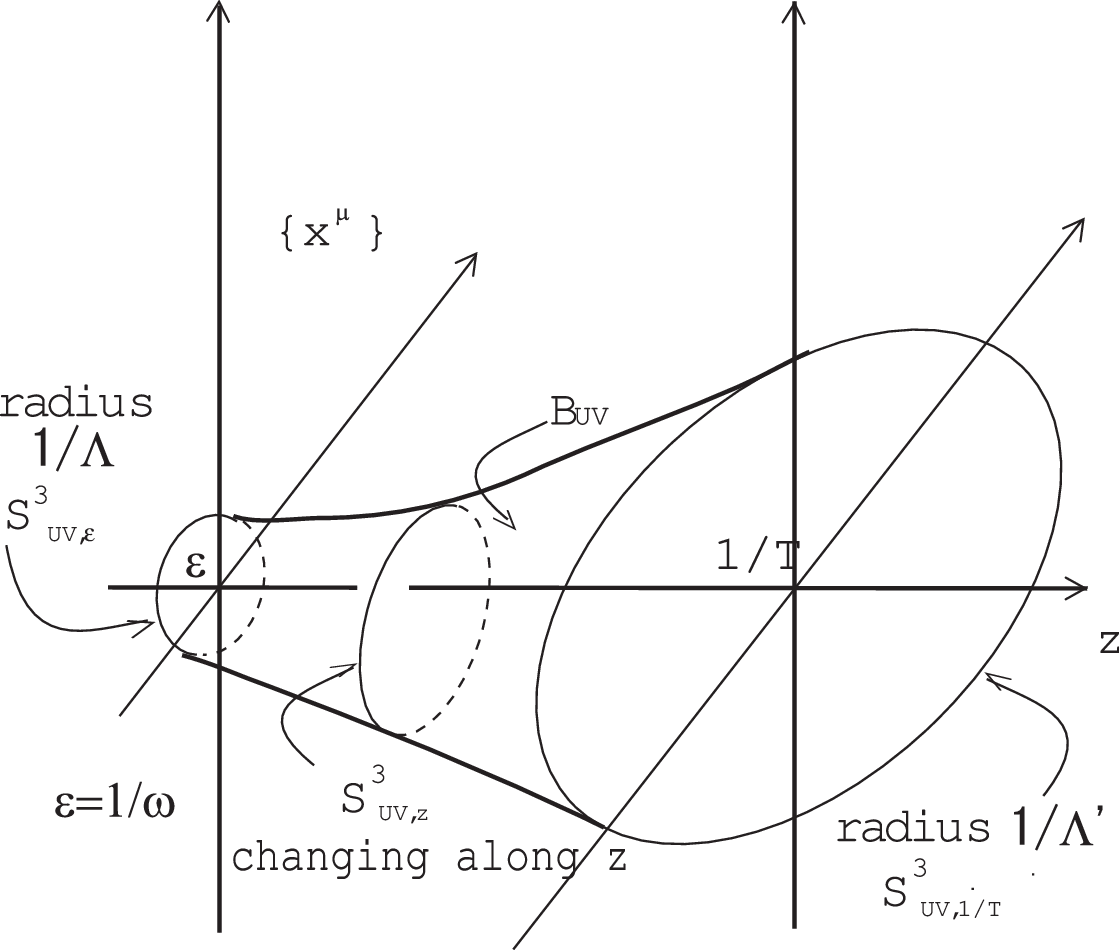}
\caption{\label{UVsurfaceW}UV regularization surface ($B_{UV}$) in 5D coordinate space.} 
\end{minipage}\hspace{2pc}%
\begin{minipage}{14pc}
\includegraphics[width=14pc]{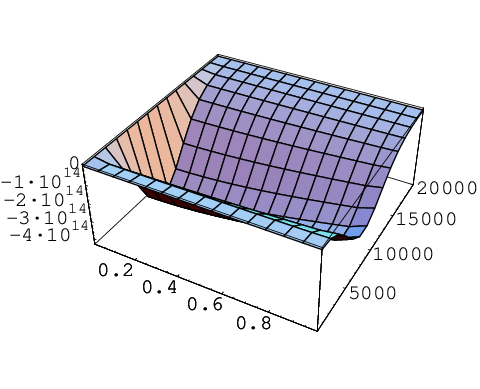}
\caption{\label{W1L2mank5senT1}Behavior of $(-N_1/2)\ptil^3W_1(\ptil,z)F^-(\ptil,z)$(elliptic suppression). 
$\La=20000,\ \om=5000,\ T=1$\ .  
$1.0001/\om\leq z\leq 0.9999/T ,\ \m=\La T/\om\leq \ptil\leq \La$.}
\end{minipage} 
\end{figure}

Although they claim the holography is behind the procedure, 
the legitimateness of the restriction looks less obvious. We have proposed 
an alternate approach 
and given a legitimate explanation within the 5D QFT\cite{IM0703,SI07Nara,SI0801,SI0803OCU}. 
Here we closely examine the {\it new regularization}. 
\begin{figure}[h]
\begin{minipage}{14pc}
\includegraphics[width=14pc]{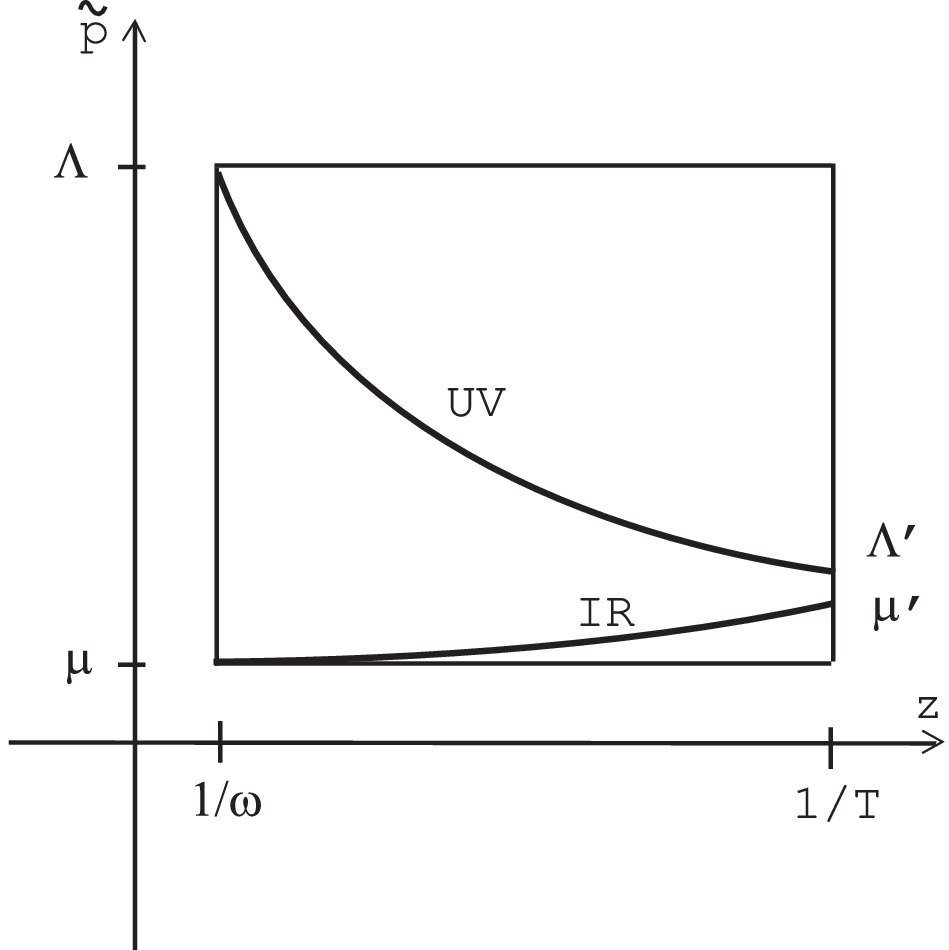}
\caption{\label{zpINTregionW2}Space of ($\ptil$,z) for the integration (present proposal).}
\end{minipage}\hspace{2pc}%
\begin{minipage}{22pc}
\includegraphics[width=22pc]{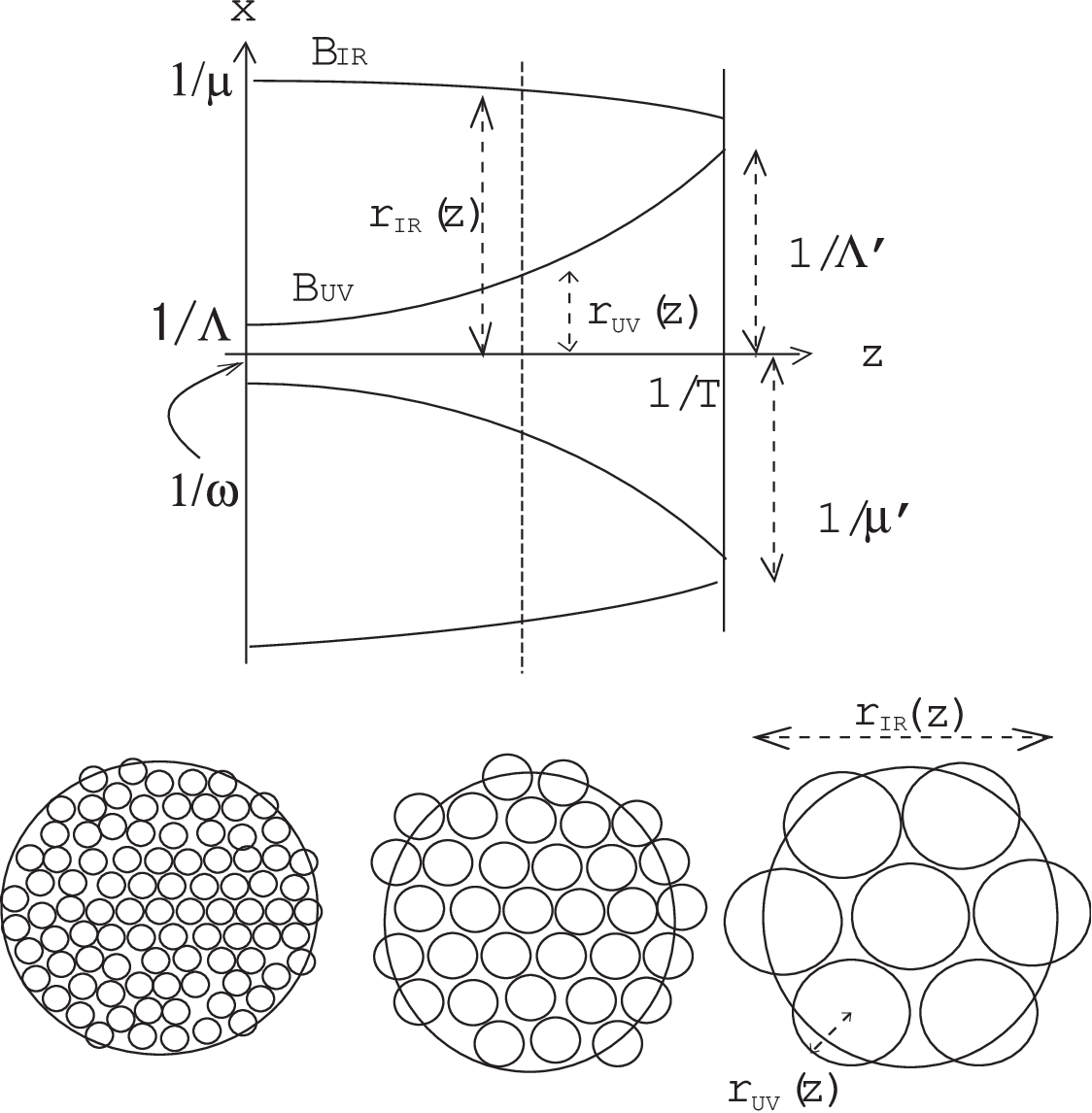}
\caption{\label{IRUVRegSurfW}Regularization Surface $B_{IR}$ and $B_{UV}$ in the 5D coordinate space $(x^\m,z)$.}
\end{minipage} 
\end{figure}
On the "3-brane" at $z=1/\om$, we introduce the IR-cutoff $\mu=\La\cdot\frac{T}{\om}$ and 
the UV-cutoff $\La$\ ($\mu\ll\La$) in the way\ :\ Eq.(5A)\   
$\mu\ \ll\ \La\ (T\ \ll\ \om)$. 
See Fig.\ref{zpINTregionW2}. 
This is legitimate in the sense that we
generally do this procedure in the 4D {\it renormalizable} thoeries. 
(Here we are considering those 5D theories that are {\it renormalizable} in "3-branes". Examples are
5D free theories (present model), 
5D electromagnetism\cite{SI0801}, 5D $\Phi^4$-theory, 5D Yang-Mills theory, e.t.c..)
In the same reason, on the
"3-brane" at $z=1/T$, we may have another set of IR and UV-cutoffs, 
$\mu'$ and $\La'$. 
We consider the case:\ Eq.(5B)\  
$\mu'\leq\La',\ \La'\ll \La,\ \mu\sim\mu'$. 
This case will lead us to
introduce the {\it renormalization flow}. (See the later discussion.)
We claim here,  
as for the regularization treatment of the "3-brane" located at other points $z$ ($1/\om<z\leq 1/T$), the regularization 
parameters are determined by the {\it minimal area principle}. 
To explain it, we move to the 5D coordinate space ($x^\m,z$). See Fig.\ref{IRUVRegSurfW}. 
The $\ptil$-expression can be replaced by $\sqrt{x_\m x^\m}$-expression by the 
{\it reciprocal relation}\ :\ Eq.(5C)\ 
$\sqrt{x_\mu(z)x^\mu(z)}\equiv r(z)\ \change \ \frac{1}{\ptil(z)}$. 
The UV and IR cutoffs change their values along $z$-axis and their trajectories make
{\it surfaces} in the 5D bulk space $(x^\mu,z)$. 
We {\it require} the two surfaces do {\it not cross} for the purpose 
of the renormalization group interpretation (discussed later).  
We call them UV and IR regularization (or boundary) surfaces($B_{UV},B_{IR}$).
The cross sections of the regularization surfaces at $z$ are the spheres $S^3$ with the 
radii $r_{UV}(z)$ and $r_{IR}(z)$. Here we consider the Euclidean space for simplicity.
The UV-surface is stereographically shown in Fig.\ref{UVsurfaceW} and reminds us of the {\it closed string} propagation. 
Note that the boundary surface B$_{UV}$ (and B$_{IR}$) is the 4 dimensional manifold. 

The 5D volume region bounded by $B_{UV}$ and $B_{IR}$ is the integral region 
 of the Casimir energy $E_{Cas}$. 
The forms of $r_{UV}(z)$ and $r_{IR}(z)$ can be
determined by the {\it minimal area principle}.
\bea
3+\frac{4}{z}r'r-\frac{r''r}{{r'}^2+1}=0
\com\q r'\equiv\frac{dr}{dz}
\com\q r''\equiv\frac{d^2r}{dz^2}
\com\q 1/\om\leq z\leq 1/T
\pr
\label{surf2}
\eea
We have confirmed, by numerically solving the above differential eqation 
(Runge-Kutta), those curves 
that show the flow of renormalization really occur. 
The results imply the {\it boundary conditions} 
determine the property of the renormalization flow.

The present regularization scheme gives the {\it renormalization group} interpretation
to the change of physical quantities along the extra axis. 
See Fig.\ref{IRUVRegSurfW}. 
\footnote{
This part is contrasting with AdS/CFT approach where 
the renormalization flow comes from the Einstein equation of 5D supergravity. 
} 
In the "3-brane" located at $z$, the UV-cutoff is $r_{UV}(z)$
and the regularization surface is the sphere $S^3$ with the radius $r_{UV}(z)$. 
The IR-cutoff is $r_{IR}(z)$
and the regularization surface is the another sphere $S^3$ with the radius $r_{IR}(z)$. 
We can regard the regularization integral region as the {\it sphere lattice} of
the following properties:\ 
a) A unit lattice (cell) is the sphere $S^3$ with radius $r_{UV}(z)$ and its inside;\ 
b) Total lattice is the sphere $S^3$ with radius $r_{IR}(z)$ and its inside;\ 
c) The integration region of this regularization is made of many cells and 
the total number of cells is const.$\times\left(\frac{r_{IR}(z)}{r_{UV}(z)} \right)^4$. 
The total number of cells changes from $(\frac{\La}{\mu})^4$ at $z=1/\om$ to 
$(\frac{\La'}{\mu'})^4$ at $z=1/T$. Along the $z$-axis, the number increases or decreases as\ :\ Eq.(5D)\ 
$\left(\frac{r_{IR}(z)}{r_{UV}(z)} \right)^4\equiv N(z)$. 
For the "scale" change $z\ra z+\Del z$, $N$ changes as\ :\ Eq.(5E)\  
$\Del (\ln N)=4\frac{\pl}{\pl z}\{\ln (\frac{r_{IR}(z)}{r_{UV}(z)}) \}\cdot \Del z$.\ 
When the system has some coupling $g(z)$, 
the renormalization group ${\tilde \be}(g)$-function (along the extra axis) is expressed as
\bea
{\tilde \be}=\frac{\Del(\ln g)}{\Del(\ln N)}=\frac{1}{\Del(\ln N)}\frac{\Del g}{g}
=\frac{1}{4}\frac{1}{\frac{\pl}{\pl z}\ln(\frac{r_{IR}(z)}{r_{UV}(z)})}\frac{1}{g}\frac{\pl g}{\pl z}
\com
\label{surf6}
\eea
where $g(z)$ is a renormalized coupling at $z$. 
\footnote{
Here we consider an interacting theory, such as 5D Yang-Mills theory and 
5D $\Phi^4$ theory, where the coupling $g(z)$ is the renormalized one 
in the '3-brane' at $z$.
}

We have explained, in this section, that the {\it minimal area principle} determines the 
flow of the regularization surfaces.

\section{Weight Function and Casimir Energy Evaluation\label{uncert}}
In the expression (\ref{HKA11}), the Casimir energy is written by
the integral in the ($\ptil,z$)-space over the range: 
$1/\om\leq z\leq 1/T,\ 0\leq \ptil\leq\infty$. In Sec.\ref{surf}, 
we have seen {\it the integral region should be properly restricted} because the cut-off
region in the 4D world 
{\it changes along the extra-axis} obeying the bulk (warped) geometry 
({\it minimal area principle}). 
We can expect the singular behavior (UV divergences) reduces by the integral-region 
restriction, but the concrete evaluation along the proposed prescription is practically not easy. 
In this section, we consider an alternate approach which respects 
the {\it minimal area principle} and evaluate the Casimir energy. 

We introduce, instead of restricting the integral region, 
a {\it weight function} $W(\ptil,z)$ in the ($\ptil,z$)-space  
for the purpose of suppressing UV and IR divergences of the Casimir Energy. 
\bea
-E^{\mp~W}_{Cas}(\om,T)\equiv\intpE\int_{1/\om}^{1/T}dz~ W(\ptil,z)F^\mp (\ptil,z)\com\q
\ptil=\sqrt{p_4^2+p_1^2+p_2^2+p_3^2}\com\nn
\mbox{Examples of}~ W(\ptil,z):\q W(\ptil,z)=\hspace{5cm}\nn
\left\{
\begin{array}{cc}
(N_1)^{-1}\e^{-(1/2) \ptil^2/\om^2-(1/2) z^2 T^2}\equiv W_1(\ptil,z),\ N_1=1.711/8\pi^2 & \mbox{elliptic suppr.}\\
(N_{2})^{-1}\e^{-\ptil zT/\om}\equiv W_2(\ptil,z),\ N_2=2\frac{\om^3}{T^3}/8\pi^2                   & \mbox{hyperbolic suppr.1}\\
(N_{8})^{-1}\e^{-1/2 (\ptil^2/\om^2+1/z^2T^2)}\equiv W_8(\ptil,z),\ N_8=0.4177/8\pi^2 & \mbox{reciprocal suppr.1}\\
\end{array}
           \right.
\label{uncert1}
\eea
where 
$F^\mp(\ptil,z)$ are defined in (\ref{HKA13}). 
In the above, 
we list some examples expected for the weight function $W(\ptil,z)$. 
$W_2$ is regarded to correspond to the regularization taken by
Randall-Schwartz. 
How to specify the form of $W$ is the subject of the next section. 
We show the shape of the energy integrand $(-1/2)\ptil^3W_1(\ptil,z)F^-(\ptil,z)$ in 
Fig.\ref{W1L2mank5senT1}. 
We notice the valley-bottom line $\ptil\approx 0.75\La$, which appeared in the un-weighted 
case (Fig.\ref{p3FmL10000}), is replaced by a new line:\ 
$\ptil^2+z^2\times \om^2T^2\approx \mbox{const}$. 
It is located {\it away from} the original $\La$-effected line
($\ptil\sim 0.75\La$).

We can check the divergence (scaling) behavior of $E^{\mp~W}_{Cas}$ by 
{\it numerically} evaluating the $(\ptil,z)$-integral (\ref{uncert1}) for 
the rectangle region of Fig.\ref{zpINTregionW}. 
\bea
-E^W_{Cas}=
\left\{
\begin{array}{cc}
\frac{\om^4}{T}\La\times 1.2\left\{  1+0.11~\ln\frac{\La}{\om}-0.10~ \ln\frac{\La}{T}  \right\} & \mbox{for}\q W_1 \\
\frac{T^2}{\om^2}\La^4\times 0.062\left\{  1+0.03~\ln\frac{\La}{\om}-0.08~ \ln\frac{\La}{T}  \right\}   &\mbox{for}\q W_2\\
\frac{\om^4}{T}\La\times 1.6\left\{  1+0.09~\ln\frac{\La}{\om}-0.10~ \ln\frac{\La}{T}  \right\}            & \mbox{for}\q W_8
\end{array}
           \right.
\label{uncert1bX}
\eea
The suppression behavior of $W_2$ improves, compared with (\ref{surfM1}) by 
Randall-Schwartz. The quintic divergence of (\ref{surfM1}) reduces to the quartic divergence
in the present approach of $W_2$. 
The hyperbolic suppressions, however, are still insufficient for the renormalizability. 
After dividing by the normalization factor, $\La T^{-1}$, 
the cubic divergence remains. 
The desired cases are others. 
The Casimir energy for each case consists of three terms. 
The first terms give {\it finite} values after dividing by the overall normalization 
factor $\La T^{-1}$. The last two terms are proportional to $\log\La$ and show 
the {\it anomalous scaling}. Their contributions are order of $10^{-1}$ to the first leading terms. 
The second ones ($\ln\frac{\La}{\om}$) contribute positively while 
the third ones ($\ln\frac{\La}{T}$) negatively.   
They give, after normalizing the factor $\La/T$, {\it only} the {\it log-divergence}. 
\bea
E^W_{Cas}/\La T^{-1} =-\al \om^4\left( 1-4c\ln (\La/\om) -4c'\ln (\La/T) \right) 
\com
\label{uncert1c}
\eea
where $\al, c$ and $c'$ can be read from (\ref{uncert1bX}) depending on the choice of $W$. 
This means the 5D Casimir energy is {\it finitely} obtained by the ordinary 
renormalization of the warp factor $\om$. (See the final section.)
In the above result of the warped case, the IR parameter $l$ in the flat result (\ref{uncert1cc}) 
is replaced by the inverse of the warp factor $\om$. 

So far as the legitimate reason of the introduction of $W(\ptil,y)$ is not clear, 
we should regard this procedure as a {\it regularization} to 
define the higher dimensional theories. 
We give a clear definition of $W(\ptil,y)$ and a legitimate explanation
in the next section. 
It should be done, in principle, in a consistent way with the bulk geometry and the gauge
principle.

\section{Meaning of Weight Function and Quantum Fluctuation of Coordinates and Momenta\label{weight}}

In the previous work\cite{SI0801}, we have presented the following idea to 
define the weight function $W(\ptil,z)$. In the evaluation (\ref{uncert1}): 
\bea
-E^{W}_{Cas}(\om,T)=\intpE\int_{1/\om}^{1/T}dz~ W(\ptil,z)F^\mp (\ptil,z)
=\frac{2\pi^2}{(2\pi)^4}\int d\ptil\int_{1/\om}^{1/T}dz~\ptil^3 W(\ptil,z)F^\mp (\ptil,z)\ ,
\label{weight1}
\eea
the $(\ptil,z)$-integral is over the rectangle region shown in Fig.\ref{zpINTregionW2} 
(with $\La\ra\infty$ and $\m\ra 0$). $F^\mp (\ptil,z)$ is explicitly given in (\ref{HKA13}). 
Following Feynman\cite{Fey72}, 
we can replace the integral by the summation over all possible pathes $\ptil(z)$. 
\bea
-E^{W}_{Cas}(\om,T)=\int\Dcal\ptil(z)\int_{1/\om}^{1/T}dz~S[\ptil(z),z]\ ,\ 
S[\ptil(z),z]=\frac{2\pi^2}{(2\pi)^4}\ptil(z)^3 W(\ptil(z),z)F^\mp (\ptil(z),z).
\label{weight1a}
\eea
There exists the {\it dominant path} $\ptil_W(z)$ which is 
determined by the minimal principle
: 
$\del S=0$.
\bea
\mbox{Dominant Path }\ptil_W(z)\ :\ \q
\frac{d\ptil}{dz}=\frac{-\frac{\pl\ln(WF)}{\pl z}}{\frac{3}{\ptil}+\frac{\pl\ln (WF)}{\pl\ptil}}
\pr
\label{weight1b}
\eea
Hence it is fixed by $W(\ptil,z)$. An example is the valley-bottom line in Fig.\ref{W1L2mank5senT1}. 
On the other hand, there exists another independent path: the minimal surface  
curve $r_g(z)$. 
\bea
\mbox{Minimal Surface Curve }r_g(z)\ :\q
3+\frac{4}{z}r'r-\frac{r''r}{{r'}^2+1}=0\com\q
\frac{1}{\om}\leq z\leq \frac{1}{T}
\com\label{weight2}
\eea 
which is obtained by the {\it minimal area principle}:\ $\del A=0$ where 
\bea
ds^2=(\del_{ab}+\frac{x^ax^b}{(rr')^2} )
\frac{dx^a dx^b}{\om^2 z^2}\equiv g_{ab}(x)dx^adx^b ,
A=\int\sqrt{\det g_{ab}}~d^4x
=\int_{1/\om}^{1/T}\frac{1}{\om^4z^4}\sqrt{{r'}^2+1}~r^3 dz.
\label{weight3}
\eea 
Hence $r_g(z)$ is fixed by the {\it induced geometry} $g_{ab}(x)$. 
Here we put the {\it requirement}\cite{SI0801}:\ Eq.(7A)\ 
$\ptil_W(z)=\ptil_g(z)$, 
where $\ptil_g\equiv 1/r_g$. This means the following things. 
We {\it require} 
the dominant path coincides with the minimal surface line $\ptil_g(z)=1/r_g(z)$ which is 
defined independently of $W(\ptil,z)$. 
In other words, $W(\ptil,z)$ is defined here by 
the induced geometry $g_{ab}(x)$. 
In this way, we can connect the integral-measure over the 5D-space with the (bulk) geometry. 
We have confirmed the (approximate) coincidence by the 
numerical method. 

In order to most naturally accomplish 
the above requirement, we can go to a 
{\it new step}. Namely, 
we {\it propose} to {\it replace} the 5D space integral with the weight $W$, (\ref{weight1}), 
by the following {\it path-integral}. We 
{\it newly define} the Casimir energy in the higher-dimensional theory as follows.  
\bea
-\Ecal_{Cas}(\om,T,\La)\equiv 
\int_{1/\La}^{1/\m}d\rho\int_{
\begin{array}{l}
r(1/\om)\\
=r(1/T)\\
=\rho
\end{array}                  }
\prod_{a,z}\Dcal x^a(z)F(\frac{1}{r},z)
~\exp\left[ 
-\frac{1}{2\al'}\int_{1/\om}^{1/T}\frac{1}{\om^4z^4}\sqrt{{r'}^2+1}~r^3 dz
    \right],
\label{weight5}
\eea 
where $\m=\La T/\om$ and the limit $\La T^{-1}\ra \infty$ is taken. 
The string (surface) tension parameter $1/2\al'$ is introduced.
 (Note: Dimension of $\al'$ is [Length]$^4$. ) 
The square-bracket ($[ \cdots ]$)-parts of (\ref{weight5}) are \ 
$-\frac{1}{2\al'}$Area = $-\frac{1}{2\al'}\int\sqrt{\mbox{det}g_{ab}}d^4x$ 
where $g_{ab}$ is the induced metric on the 4D surface. 
$F(\ptil,z)$ is defined in (\ref{uncert1}) or (\ref{HKA13}) and shows 
the {\it field-quantization} of the bulk scalar (EM) fields. 
In the above expression, we have followed 
the path-integral formulation of the {\it density matrix} (See Feynman's text\cite{Fey72}). 
The validity of the above definition is based on the following points: 
a)\ When the weight part (exp $[\cdots]$-part) is 1, the proposed quantity 
$\Ecal_{Cas}$ is equal to $E^W_{Cas}$, 
(\ref{weight1}), with $W=1$ 
;\ 
b)\ The leading path is given by $r_g(z) =1/p_g(z)$, (\ref{weight2});\ 
c)\ The proposed definition, (\ref{weight5}), clearly shows the 4D space-coordinates $x^a$ 
or the 4D momentum-coordinates $p^a$ are {\it quantized} (quantum-statistically, not field-theoretically) with 
the Euclidean time $z$ and the "{\it area} Hamiltonian" 
$A=\int\sqrt{\det g_{ab}}~d^4x$. Note that $F(\ptil,z)$ or $F(1/r,z)$ 
appears, in (\ref{weight5}), as the {\it energy density operator} in the quantum statistical system of
$\{ p^a(z)\}$ or $\{ x^a(z)\}$. 

In the view of the previous paragraph, the treatment of Sec.\ref{uncert} is an {\it effective} action 
approach using the (trial) weight function $W(\ptil,z)$. 
Note that the integral over $(p^\m,z)$-space, appearing in (\ref{HKA13}), 
is the summation over all degrees of freedom of the 5D space(-time) points using the "naive" measure 
$d^4pdz$. 
An important point is that we have the possibility to take another  
measure for the summation in the case of the higher dimensional QFT. 
We have adopted, in Sec.\ref{uncert}, the new measure $W(p^\m,z)d^4pdz$ in such a way that the Casimir energy 
{\it does not show physical divergences}. 
We expect the {\it direct} evaluation of (\ref{weight5}), numerically 
or analytically, leads to the similar result. 
\section{Discussion and Conclusion\label{conc}}

The log-divergence in (\ref{uncert1c}) is the familiar one 
in the ordinary QFT. It can be {\it renormalized} in the 
following way. 
\bea
\frac{E^W_{Cas}}{\La T^{-1}} =-\al\om^4\left( 1-4c\ln (\frac{\La}{\om})-4c'\ln (\frac{\La}{T}) \right) 
=-\al (\om_r)^4, 
\om_r=\om\sqrt[4]{1-4c\ln (\frac{\La}{\om})-4c'\ln (\frac{\La}{T}) },
\label{conc1}
\eea
where $\om_r$ is the {\it renormalized} warp factor and $\om$ is the {\it bare} one. 
No local counterterms are necessary. 
Note that this renormalization relation is {\it exact} (not a perturbative result). 
In the familiar case of the 4D renormalizable theories, the coefficients $c$ and $c'$ depend on the coupling, but, 
in the present case, they are {\it pure numbers}. 
It reflects the {\it interaction between (EM) fields and the boundaries}. 
When $c$ and $c'$ are sufficiently small 
we find the renormalization group function for the warp factor $\om$ as  
\bea
|c|\ll 1\ ,\ |c'|\ll 1\com\q 
\om_r=\om (1-c\ln (\La/\om)-c'\ln (\La/T) )\ ,\ 
\be\equiv \frac{\pl}{\pl(\ln \La)}\ln \frac{\om_r}{\om}=-c-c'
\pr
\label{conc2}
\eea
We should notice that, in the flat geometry case, the IR parameter (extra-space size) $l$ 
is renormalized. In the present warped case, however, the corresponding parameter $T$ 
is {\it not renormalized}, but the {\it warp parameter} $\om$ {\it is renormalized}. 
Depending on the sign of $c+c'$, the 5D bulk curvature $\om$ 
{\it flows} as follows. 
When $c+c'>0$, the bulk curvature $\om$ decreases (increases) as the 
the measurement energy scale $\La$ increases (decreases). 
When $c+c'<0$, the flow goes in the opposite way. 
When $c+c'=0$, $\om$ does not flow ($\be=0$) and is given by 
$\om_r=\om (1+c\ln(\om/T))$. 

The final result (\ref{conc1}) is the {\it new} type Casimir energy, $-\om^4$. 
$\om$ appears as a boundary parameter like $T$. 
The familiar one 
is $-T^4$ in the present context. 
In ref.\cite{IM05NPB}, 
another type $T^2\om^2$ was predicted using a "quasi" Warped model (bulk-boundary theory). 

Through the Casimir energy calculation, in the higher dimension, we find a way to 
quantize the higher dimensional theories within the QFT framework. 
The quantization {\it with respect to the fields} (except the gravitational fields $G_{AB}(X)$) 
is done in the standard way. After this step, the expression has the summation 
{\it over the 5D space(-time) coordinates or momenta} 
$\int dz\prod_adp^a$. We have proposed that this summation should be replaced by 
the {\it path-integral} $\int \prod_{a,z}\Dcal p^a(z)$ with the {\it area} action (Hamiltonian) 
$A=\int\sqrt{\det g_{ab}}d^4x$ where $g_{ab}$ is the {\it induced} metric on the 4D surface. 
This procedure says the 4D momenta 
$p^a$ (or coordinates $x^a$) are {\it quantum statistical} operators and 
the extra-coordinate $z$ is the inverse temperature (Euclidean time). 
We recall the similar situation occurs in the standard string approach. 
The space-time coordinates obey some uncertainty principle\cite{Yoneya87}.  

Recently the dark energy (as well as the dark matter) in the universe is a hot subject. 
It is well-known that the dominant candidate is the cosmological term. 
We also know the proto-type higher-dimensional theory, that is, the 5D KK theory, 
has predicted so far the {\it divergent} cosmological constant\cite{AC83}. 
This unpleasant situation has been annoying us for a long time. 
If we apply the present result, the situation drastically improves. The cosmological
constant $\la$ appears as:\ Eq.(8A)\ 
$R_\mn-\half g_\mn R-\la g_\mn =T_\mn^{matter},
S=\int d^4x \sqrt{-g}\{\frac{1}{G_N}(R+\la) \}
+\int d^4x \sqrt{-g}\{\Lcal_{matter}\}, 
g=\mbox{det}~g_{\mn}$, 
where $G_N$ is the Newton's gravitational constant, $R$ is the Riemann scalar 
curvature. 
We consider here the 3+1 dim Lorentzian space-time ($\mu,\nu=0,1,2,3$). 
The constant $\la$ observationally takes the value\ :\ Eq.(8B)\ 
$\frac{1}{G_N}\la_{obs}\sim \frac{1}{G_N{R_{cos}}^2}\sim m_\n^4\sim (10^{-3} eV)^4
,
\la_{obs}\sim \frac{1}{R_{cos}^{~2}}\sim 4\times 10^{-66}(eV)^2$,  
where $R_{cos}\sim 5\times 10^{32}\mbox{eV}^{-1}$ is the cosmological size (Hubble length), $m_\n$ is the neutrino mass.
\footnote{ 
The relation $m_\n\sim \sqrt{M_{pl}/R_{cos}}=\sqrt{1/R_{cos}\sqrt{G_N}}$, which appears  
in some extra dimension model\cite{SI0012,SI01Tohwa}, is used. The neutrino mass is, 
at least empirically, located 
at the {\it geometrical average} of two extreme ends of the mass scales in the universe.  
} 
On the other hand, we have theoretically so far\ :\ Eq.(8C)\ 
$\frac{1}{G_N}\la_{th}\sim \frac{1}{{G_N}^2}={M_{pl}}^4\sim (10^{28} eV)^4$. 
This is because the mass scale usually comes from the quantum 
gravity. (See ref.\cite{SI83NPB} for the derivation using the  
Coleman-Weinberg mechanism.) 
We have the famous huge discrepancy factor\ :\ Eq.(8D)\ 
$\frac{\la_{th}}{\la_{obs}}\sim N_{DL}^{~2}, N_{DL}\equiv M_{pl}R_{cos}\sim 6\times 10^{60}$, 
where $N_{DL}$ is the Dirac's large number\cite{PD78}. 
If we use the present result (\ref{conc1}), we can obtain a natural choice of 
$T, \om$ and $\La$
as follows. By identifying 
$T^{-4}E_{Cas}=-\al_1\La T^{-1}\om^4/T^4$ with 
$\int d^4x\sqrt{-g}(1/G_N)\la_{ob}=R_{cos}^{~2}(1/G_N)$, we 
obtain the following relation:\ Eq.(8E)\  
$N_{DL}^{~2}=R_{cos}^{~2}\frac{1}{G_N}=-\al_1 \frac{\om^4\La}{T^5}, \al_1\ :\ \mbox{some coefficient}$. 
The warped (AdS$_5$) model predicts the cosmological constant {\it negative}, 
hence we have interest only in its absolute value.
We take the following choice for $\La$ and $\om$\ :\ Eq.(8F)\ 
$\La=M_{pl}\sim 10^{19}GeV, 
\om\sim
 \frac{1}{\sqrt[4]{G_N{R_{cos}}^2}}=\sqrt{\frac{M_{pl}}{R_{cos}}}\sim m_\n\sim 10^{-3}\mbox{eV}$. 
The choice for $\La$ is accepted in that the largest known energy scale is the Planck energy. 
The choice for $\om$ comes from the experimental bound for the Newton's gravitational force. 

As shown above, we have the standpoint that the cosmological constant is mainly made from 
the Casimir energy.  
We do not yet succeed in obtaining the value $\al_1$ negatively, but
succeed in obtaining  
the finiteness of the cosmological constant and its gross absolute value. 
The smallness of the value is naturally explained by the renormalization flow as follows. 
Because we already know the warp parameter $\om$ {\it flows} (\ref{conc2}), 
the $\la_{obs}\sim 1/R_{cos}^2$ expression (8F), $\la_{obs}\propto \om^4$, says that the {\it smallness of the cosmological constant comes from 
the renormalization flow} for the non asymptotic-free case ($c+c'<0$ in (\ref{conc2})). 

The IR parameter $T$, the normalization factor $\La/T$ in (\ref{uncert1c}) and the IR cutoff 
$\mu=\La\frac{T}{\om}$ are given by\ :\ Eq.(8G)\  
$T=R_{cos}^{~-1}(N_{DL})^{1/5}\sim 10^{-20}eV, 
\frac{\La}{T}=(N_{DL})^{4/5}\sim 10^{50}, 
\mu=M_{pl}N_{DL}^{-3/10}\sim 1GeV\sim m_N$, 
where $m_N$ is the nucleon mass. 
The Fig.\ref{IRUVRegSurfW} strongly suggests that 
the degree of freedom of the universe (space-time) 
is given by\ :\ Eq.(8H)\  
$\frac{\La^4}{\m^4}=\frac{\om^4}{T^4}=N_{DL}^{~6/5}\sim 10^{74}\sim (\frac{M_{pl}}{m_N})^4$. 

\end{document}